\documentclass[12pt]{article}
\pdfoutput=1
\usepackage{amsbsy}
\usepackage{amsfonts}
\usepackage{amsmath,amssymb}
\usepackage{bbold}
\usepackage{graphicx}

\usepackage[makeroom]{cancel}

\usepackage{color}

\newcommand{\sect}[1]{\setcounter{equation}{0}\section{#1}}

\newcommand{\app}[1]{\setcounter{section}{0}
\setcounter{equation}{0} \renewcommand{\thesection}{\Alph{section}}
\section{#1}}
\newcommand{\eq}{\begin{equation}}
\newcommand{\eqa}{\begin{eqnarray}}  
\newcommand{\en}{\end{equation}}
\newcommand{\ena}{\end{eqnarray}}
\newcommand{\enn}{\nonumber \end{equation}}

\def\sk{\vskip .4cm}
\def\noi{\noindent}

\def\al{\alpha}
\def\be{\beta}

\let \part\partial

\def\part{\partial}

\def\sk{\vskip .4cm}

\def\noi{\noindent}

\def\X0{X^0}

\def\al{\alpha}

\def\square{{\,\lower0.9pt\vbox{\hrule \hbox{\vrule height 0.2 cm
\hskip 0.2 cm \vrule height 0.2 cm}\hrule}\,}}

\def\lb{\langle}
\def\rb{\rangle}

\def\Afat{\mathbb{A}}
\def\Bfat{\mathbb{B}}
\def\Pfat{\mathbb{P}}

\def\noali{\al_i \mkern-15mu/}

\begin{document}

\begin{titlepage}
%\rightline{ARC-20-06}

\vskip 2em
\begin{center}
{\Large \bf Space and time correlations in quantum histories } \\[3em]

\vskip 0.5cm

{\bf
Leonardo Castellani${}^{1,2,3}$ and Anna Gabetti${}^4$}
\medskip

\vskip 0.5cm

{\sl ${}^1$ Dipartimento di Scienze e Innovazione Tecnologica
\\Universit\`a del Piemonte Orientale, viale T. Michel 11, 15121 Alessandria, Italy\\ [.5em] ${}^2$ INFN, Sezione di 
Torino, via P. Giuria 1, 10125 Torino, Italy\\ [.5em]
${}^3$ Arnold-Regge Center, via P. Giuria 1, 10125 Torino, Italy \\ [.5em]
${}^4$ Dipartimento di Scienza Applicata e Tecnologia, Politecnico di Torino\\
Corso Duca degli Abruzzi 24, 10129 Torino, Italy
}\\ [4em]
\end{center}

\begin{abstract}
\sk

The formalism of generalized quantum histories allows a symmetrical treatment of space and time correlations, by taking different traces of the same history density matrix. We recall how to characterize spatial and temporal entanglement in this framework. An operative protocol is presented, to map a history state into the ket of a static composite system. We 
show, by examples, how the Leggett-Garg and the temporal CHSH inequalities can be violated in our approach.

\end{abstract}

\vskip 8cm\
 \noi \hrule \vskip .2cm \noi {\small
leonardo.castellani@uniupo.it, anna.gabetti@polito.it}

\end{titlepage}

\newpage
\setcounter{page}{1}

\tableofcontents

%%%%%%%%%%%%%%%
\sect{Introduction}
%%%%%%%%%%%%%%%

The evolution of a classical system can be described by a temporal sequence (continuous or discrete)
of physical states. The states can be characterized by definite
values $\alpha_i$ of physical quantities at time $t_i$, so that the sequence $\alpha _1,...\alpha_n$ represents the time evolution of the system. We can call this sequence a {\sl history}
of the system. If the equations of motion for $\alpha(t)$ are of first order in time, the initial state $\alpha_1$ determines all successive states \footnote{In fact, a state at {\sl any} time determines all the others, past and future.}. For example a trajectory in phase space represents the history of
a classical system, governed by the Hamilton equations. In this case the history $(q_1,p_1),...,(q_n,p_n)$ is determined by the values of position and momentum at a given time. Also nonphysical trajectories, i.e. trajectories that do not satisfy the classical equations of motion, can be considered, and in fact enter in the formulation of variational principles. 

The evolution of quantum systems, in contradistinction, cannot be modeled simply by a temporal series of quantum states $|\psi(t_1)\rb,...,|\psi(t_n)\rb$, if we insist on characterizing states by definite values $\alpha_i$ of observables $A_i$ at each time $t_i$. Indeed these definite values can only be acquired by a measurement, implying in general a nondeterministic collapse of the quantum state. As a consequence the evolution of the system becomes stochastic: an initial state $|\psi\rb$ does not determine a single sequence of measurement results $\alpha _1,...\alpha_n$ (a ``history"), but a whole tree of quantum histories, cf. Fig. 1. Since probabilities of measuring an outcome $\alpha_i$ only depend on $\alpha_{i-1}$ 
({\sl and} on the evolution operator between times $t_{i-1}$ and $t_i$), Fig. 1 describes a sort of ``Markov" tree.

Since the work of Feynman \cite{Feynman,FH} (see also Dirac \cite{Dirac}), this idea has led to various formulations based on histories, rather than on states at a given time. A very partial list of references, relevant for the present paper, is given in \cite{histories1} - \cite{histories15}. 

Essentially in all approaches, histories are defined by sequences of measurements of observables at different times $t_1,...t_n$ (observables that can be chosen differently at each time), generalizing the histories of positions in configuration space considered in the Feynman path integral.

~~~~~~~~~~~~~~~~~~\includegraphics[scale=0.30]{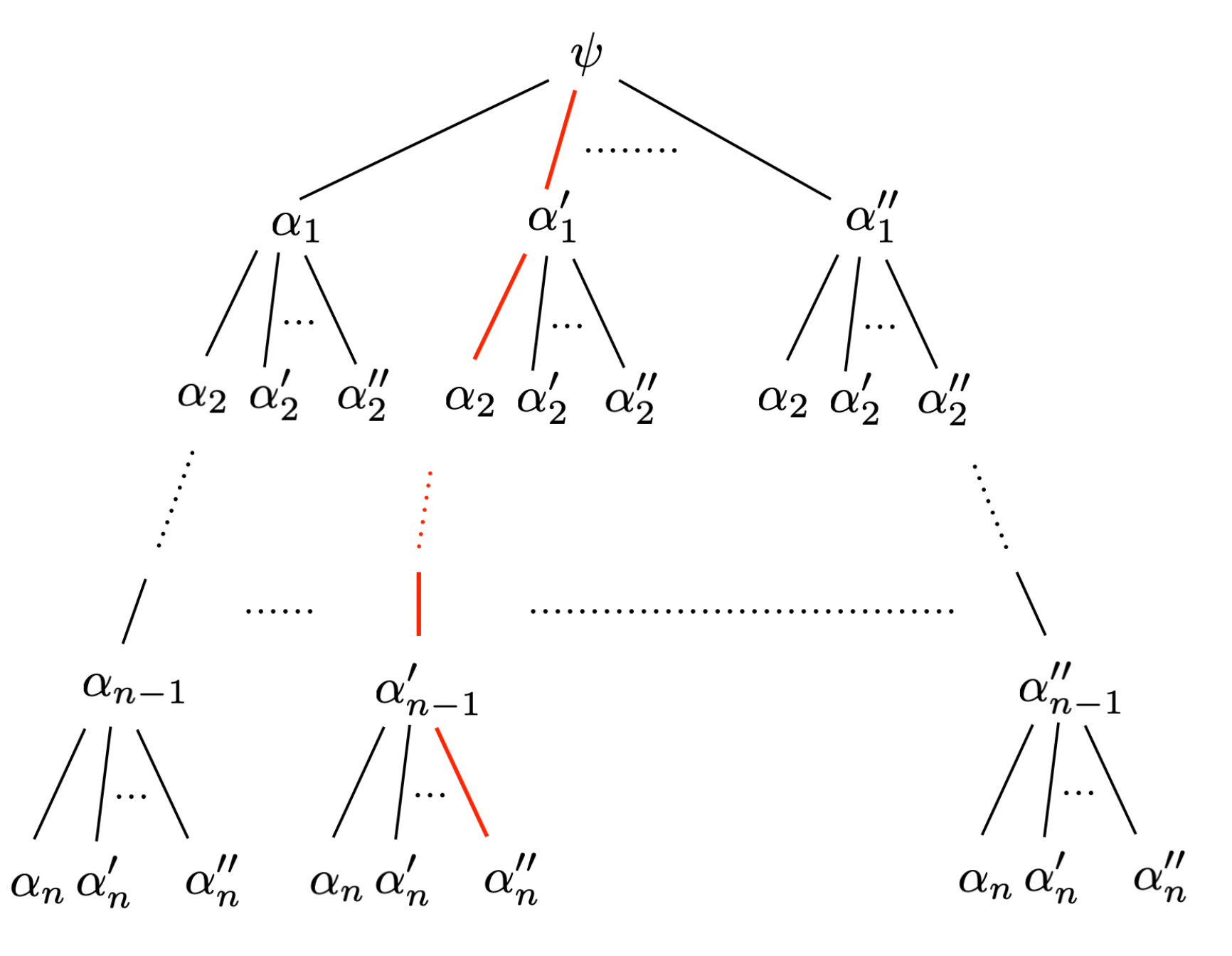}

~~~~~~~~~~~~~~~~~~~~~~~\noi {\bf Fig. 1 } {\small The tree of quantum histories}

\sk
In the present paper we summarize the formalism of history vectors, developed in ref.s \cite{LC1,LC2,LC3}, and discuss some new applications of it.

In this formalism, an evolving system with initial state $\psi$ at time $t_0$ is described by a vector living in a tensor product  ${\cal H}_1 \odot \cdots \odot {\cal H}_n$, where ${\cal H}_i$ is the Hilbert space of the system at time $t_i$. As is customary in history approaches to quantum mechanics, we use the symbol $\odot$ for the ``temporal" tensor product, to distinguish it from the usual $\otimes$ product between subsystems of a composite system. The history vector is a superposition of all the basis vectors $|\alpha_1\rb \odot \cdots \odot |\alpha_n\rb$, where $|\alpha_i\rb$ are the eigenvectors (with eigenvalues $\alpha_i$) of observables $A_i$, measured at each time $t_i$. The dynamical information is contained in the coefficients of the superposition, which are the history amplitudes $A(\alpha_1,...\alpha_n)$, whose
square modulus gives the probability of obtaining the sequence $\alpha_1,...\alpha_n$ in measuring $A_1,...A_n$ at times $t_1,...t_n$. This is a direct generalization of the expansion of a state $|\psi(t_1)\rb$, describing a system at time $t_1$, having started in the state $|\psi\rb$ at $t_0$:
\eq
|\psi(t_1)\rb = \sum_{\alpha_1} A(\psi,\alpha_1) |\alpha_1\rb
\en
where the amplitude is $A(\psi,\alpha_1)=\lb \alpha_1 |U(t_1,t_0) |\psi\rb$ and $U(t_1,t_0)$ is the evolution operator between $t_0$ and $t_1$.

An operative protocol is proposed, to construct a state of a composite system that contains the same amplitudes of the history vector, thereby exchanging evolution with compositeness. On this state one can easily test
entanglement, signalling in fact temporal entanglenent in the original system. The definitions of density matrices,
entangled history states, von Neumann entropy, and temporal entanglement entropy are particularly simple in our 
approach. Comparing with the consistent (or decoherent) histories approach of ref.s \cite{histories1,histories2,histories2bis}, we show that the Leggett-Grag inequality can never be violated in that framework, whereas it is very easy to find examples of its violation using history vectors, containing all histories with nonvanishing amplitudes. Finally, examples of the temporal CHSH violation are discussed.

The paper is organized as follows. In Section 2 we recall how to compute probabilities of outcome sequences, and
in Section 3 we introduce history vectors. Section 4 deals with history observables, and Section 5 is a resumé of
history density matrices and their space and time reduction, allowing to define space and time entanglement entropy in 
a completely symmetric way. A protocol to construct 
a (static) composite system that models time evolution is presented in Section 6. In Section 7 we demonstrate that
considering only consistent histories can never lead to violations of the Leggett-Garg inequalities.
Finally in Section 8 we discuss the temporal CHSH inequality within the history vector framework, and
provide examples of its violation. Section 9 contains some conclusions.

%%%%%%%%%%%%%%
\sect{History probabilities}
%%%%%%%%%%%%%%

Each path in the tree of Fig.1 can be assigned a probability $p(\alpha)$, i.e. the probability obtaining the results $\alpha _1,...\alpha_n$ in a sequence of measurements on the system at times $t_1,...t_n$:
\eq
p(\alpha)= |A(\psi,\alpha)|^2   \label{prob1}
\en
where $A(\psi,\alpha)$ is the {\sl amplitude} of the particular quantum history, given by
\eq
A(\psi,\alpha) = \lb \alpha_n |U(t_n,t_{n-1}) P_{\alpha_{n-1}} U(t_{n-1},t_{n-2}) P_{\alpha_{n-2}} \cdots P_{\alpha_1} U(t_1,t_0) |\psi \rb \label{amplitude}
\en
The unitary operator $U(t_i, t_{i-1})$ is the evolution operator from time $t_{i-1}$ to time $t_i$, and $P_{\alpha_i}$ are the projectors on the eigensubspaces corresponding to the eigenvalues $\alpha_i$,
satisfying the ortogonality and completeness relations
\eq
P_{\alpha_i} P_{\alpha'_i} = P_{\alpha_i}  \delta_{\alpha_i,\alpha'_i},~~~\sum_{\alpha_i} P_{\alpha_i} = I
\label{projectorprop}
\en
The probability (\ref{prob1}) to obtain the sequence $\alpha _1,...\alpha_n$, is obtained by repeated application of the Born rule. For simplicity in the following we suppose that these eigenvalues completely characterize the state at time $t_i$, i.e. they are eigenvalues of a complete set of commuting observables that are measured at time $t_i$. Therefore 
\eq
P_{\alpha_i}  = | \alpha_i \rb \lb \alpha_i |
\en
Thus if the system initial state $|\psi\rb$ is known, together with its Hamiltonian, we can 
calculate the amplitudes and probabilities of the sequences of measurement results corresponding to chosen observables at each time $t_i$. This is the starting point of several ``history formulations" of
quantum mechanics, including the one we use in the present paper, first proposed in \cite{LC2}.

Notice that, from unitarity of $U$ and the projector properties (\ref{projectorprop}), the amplitudes satisfy:
\eq
 \sum_{\alpha_i} A(\psi,\al_1,\cdots, \al_i,\cdots, \al_n)  = A(\psi,\al_1,\cdots, \noali~~,\cdots \al_n) \label{sumruleA} 
 \en
 History probabilities as computed in (\ref{prob1}) satisfy
 \eq
 \sum_\al p(\al)=1
 \en
 as expected, but satisfy the marginal sum rules only in the form
\eq
 \sum_{\alpha_n} p(\al_1,\cdots, \al_n)  = p(\al_1,\cdots, \cdots \al_{n-1}) \label{sumrulep} 
 \en
i.e. only when the sum involves the last eigenvalue $\al_n$. This implies also
\eq
 \sum_{\alpha_{i+1},...\alpha_n} p(\al_1,\cdots, \al_i, \al_{i+1},\cdots \al_n)  = p(\al_1,\cdots \al_{i}) \label{sumrulep2} 
 \en
 For this reason many authors speak of history {\sl weights} rather than probabilities. But 
 the $p(\alpha)$ as computed in (\ref{prob1}) give indeed the {\sl probabilities} of obtaining the sequences
 $\alpha$ in successive measurements on the quantum system, even if these probabilities do not
 satisfy all the classical sum rules. 
 
 This discrepancy is the signature of a quantum system: since a measurement in general modifies its state, while in classical systems the state is unaffected by an ideal measurement, differences in sum rules on intermediate results are to be expected.    
 \sk
 \noi {\bf Note:} amplitudes and probabilities can be given in terms of {\sl chain operators}:
\eq
C_{\psi,\al} = P_{\alpha_{n}}  U(t_n,t_{n-1}) ~ P_{\alpha_{n-1}} ~ U(t_{n-1},t_{n-2})  \cdots P_{\alpha_{1}}~
U(t_1,t_0) P_{\psi} \label{chain}
\en
with $P_\psi=|\psi\rb\lb \psi |$. Indeed:
\eqa
& & C_{\psi,\al}=|\al_n\rb A(\psi,\al) \lb \psi | \\
& & p(\psi,\al)= |A(\psi,\al)|^2 = Tr(C_{\psi,\al} C^\dagger_{\psi,\al}) 
\ena
It is not difficult to prove that probabilities satisfy {\sl all} the classical marginal rules:
\eq
 \sum_{\alpha_i} p(\psi,\al_1,\cdots, \al_i,\cdots, \al_n)  = p(\psi,\al_1,\cdots, \noali~~,\cdots \al_n) \label{sumrulepC} 
 \en
 if and only if the so-called {\it decoherence condition} is satisfied:
\eq
Tr (C_{\psi,\al} C_{\psi,\be}^\dagger) + c.c.= 0~~when~\al \not= \be  \label{decocondition}
\en
If all the histories we consider are such that the decoherence condition holds, they are said to form a {\it consistent} (or {\sl decoherent}) set, and can be assigned probabilities satisfying all the standard sum rules.

In general, histories do not form a consistent set: interference effects between them can be important, as in the case of the double slit experiment. For this reason we will not limit ourselves to consistent sets. Formula (\ref{prob1}) for the probability of successive measurement outcomes holds true in any case.

 %%%%%%%%%%%%%%%%%%%%
 \sect{History vectors}
 %%%%%%%%%%%%%%%%%%%%

Can we generalize the ket description of a quantum system to a description that 
encodes its evolution, including measurement and collapse ? 

Suppose that the state is $|\psi (t_0) \rb= |\psi \rb$ at time $t_0$, that 
its evolution at time $t_1$ 
is $|\psi(t_1)\rb = U(t_1,t_0) |\psi (t_0) \rb$, and that we have
devices to measure a complete set of commuting observables at time $t_1$, with eigenvalues $\alpha_1$ and corresponding eigenvectors $|\alpha_1\rb$.
Then using $I=\sum_{\al_1} |\al_1\rb \lb \al_1 |$, the state $|\psi (t_1) \rb$ can be 
expanded as
\eq
|\psi (t_1) \rb = \sum_{\alpha_1}  |\alpha_1 \rb \lb \alpha_1 |U(t_1,t_0) |\psi \rb =
\sum_{\alpha_1} A(\psi,\alpha_1) |\alpha_1 \rb
\en
We can describe the quantum system and its evolution via a ``temporal" tensor product
\eq
| \Psi \rb = |\psi (t_0)\rb \odot |\psi (t_1)\rb = \sum_{\alpha_1} A(\psi, \alpha_1) |\psi \rb \odot |\alpha_1\rb
\label{historyvector01}
\en
where linearity of the tensor product has been used. In other words, we describe the system as a linear combination 
of the histories $\psi, \alpha_1$, where $\psi$ refers to the initial state at $t_0$ and $\alpha_1$ is one of the
possible outcomes of a measurement at $t_1$. We have thereby replaced the ket $|\psi\rb$, describing the state of the system at time $t_0$, with a ``history ket" living in a tensor product space, containing
the information on time evolution $t_0 \rightarrow t_1$ of $|\psi\rb$.

Similarly, take now as initial state $|\alpha_1 \rb$, the state in which the system collapses after $\alpha_1$ has been obtained in a measurement. Applying the same procedure, this state can be replaced by
the history state
\eq
\sum_{\alpha_2} A(\alpha_1,\alpha_2) |\alpha_1\rb \odot |\alpha_2\rb  \label{historyalpha1}
\en
where $A(\alpha_1,\alpha_2)=\lb \alpha_2 | U(t_2,t_1) |\alpha_1 \rb$, and $\alpha_2$ are possible measurement results at time $t_2$. Inserting (\ref{historyalpha1})
in place of $|\alpha_1\rb$ in expression (\ref{historyvector01}) produces
\eq
|\Psi \rb = \sum_{\alpha_1,\alpha_2} A(\psi,\alpha_1) A(\alpha_1,\alpha_2)~ |\psi\rb \odot |\alpha_1\rb \odot |\alpha_2\rb  =  \sum_{\alpha_1,\alpha_2} A(\psi,\alpha_1,\alpha_2) ~|\psi\rb \odot |\alpha_1\rb \odot |\alpha_2\rb
\en
where we have used again linearity of the tensor product and the multiplication rule of amplitudes:
\eq
A(\psi, \alpha_1,...\alpha_i) A(\alpha_i,...\alpha_n) = A(\psi, \alpha_1,...\alpha_n)
\en
Generalizing the above considerations, we are led to describe quantum systems by means a {\sl history vector} $|\Psi\rb$, living in a temporal tensor space, and defined by a linear combination of all possible histories, each multiplied by its amplitude:
\eq
|\Psi\rb = \sum_{\alpha_1,...\alpha_n} A(\psi,\alpha_1,...,\alpha_n) ~ |\psi\rb \odot |\alpha_1\rb \odot  \cdots \odot|\alpha_n\rb \label{historyvector}
\en
In the following we will often omit the initial state $|\psi\rb$ in the temporal tensor products for notational simplicity, and occasionally indicate with $\al$ the whole sequence $\alpha_1,...\alpha_n$.

Introducing the usual scalar product between tensor states, these eigenvectors are automatically orthonormal and provide a basis in history space. The history vector is normalized since
\eq
\lb \Psi | \Psi \rb= \sum_\al |A(\psi,\al)|^2 = 1
\en
The {\it history content} of the system is defined to be the set of histories $\al = \al_1,...\al_n$ contained in
$|\Psi\rb$, i.e. all histories having nonvanishing amplitudes.
\sk
Probabilities of measuring sequences $\al=\al_1,...\al_n$ are given by the familiar Born rule
\eq
p(\psi,\al)= \lb \Psi | \mathbb{P}_\al |\Psi\rb = |A(\psi,\al)|^2.   \label{probvector}
\en
with 
\eq
\mathbb{P}_\al = |\al_1\rb\lb\al_1| \odot ... \odot  |\al_n\rb\lb\al_n| 
\en
Note that formula (\ref{probvector}) holds for sequences of measurements occurring at {\sl all} times $t_1,...t_n$. This formula can be generalized to the case of partial measurements (i.e. measurements
occurring at times $t_{i_1},...t_{i_m}$, where $i_1,...i_m$ is a subset of $1,...n$) using an appropriate
projection, see ref.\cite{LC2}.
\sk
The description of an evolving quantum system, proposed in this Section, depends crucially
on what are the observables measured at each time $t_i$. In fact the history vectors of two
quantum systems, differing only by the observables measured at times $t_i$, are in general different.
For example if the initial state is the superposed qubit state $|\psi\rb = {1 \over \sqrt{2}} (|0\rb + |1\rb)$
and the evolution operator is the identity (trivial evolution), then the history vector for measurements
at $t_1$ and $t_2$ of the observable $X$ (defined by $X|0\rb = |1\rb, X|1\rb = |0\rb$) is simply
\eq
 |X=1\rb \odot |X=1\rb = {1 \over 2} (|0\rb \odot |0\rb + |0\rb \odot |1\rb + |1\rb \odot |0\rb + |1\rb \odot |1\rb)
 \label{example1}
\en
where $|X=1\rb = |\psi\rb$ is the eigenvector of $X$ corresponding to its eigenvalue $+1$. If, on the other hand, the observable $Z$ (defined by $Z|0\rb = |0\rb, Z|1\rb =- |1\rb$)  is measured at times $t_1$ and $t_2$, the corresponding history vector becomes:
\eq
{1 \over \sqrt{2}} (|0\rb \odot |0\rb  +  |1\rb \odot |1\rb)
\en
differing from (\ref{example1}) by the absence of cross-terms.

The formalism therefore recognizes that observers (i.e. measuring devices) are 
unavoidably {\sl part} of the description of an evolving quantum system. In this sense
this description could be considered {\sl relational}, i.e. related to the particular 
measurements considered at various times. It helps to resolve so-called paradoxes,
as for example the three box game proposed in ref. \cite{threebox}, by clearly
distinguishing systems subjected to different measuring apparati throughout their evolution.

 %%%%%%%%%%%%%%%%%%%%
 \sect{History observables}
 %%%%%%%%%%%%%%%%%%%%

As discussed in the previous Section, the history description includes measuring apparati, activated at all times $t_1,...t_n$, corresponding to ``time local" observables $A_1,...A_n$. These can be extended to {\sl history observables}, acting on the whole history vector space,  given by:
\eq
\mathbb{A}_1= A_1 \odot  I \cdots \odot I,~...~ \mathbb{A}_n =I \odot \cdots \odot I \odot A_n \label{historyobservable}
\en
The history vectors $|\psi\rb \odot |\alpha_1\rb \odot  \cdots \odot|\alpha_n\rb$ are eigenvectors of these observables, with eigenvalues respectively given by $\alpha_1, ...\alpha_n$. With the usual scalar product between tensor states, these eigenvectors are automatically orthonormal and provide a basis in history space.

The projectors on eigensubspaces of $\mathbb{A}_i$ are
\eq
\mathbb{P}_{\alpha_i} = I \odot \cdots \odot |\alpha_i\rb\lb \alpha_i| \odot \cdots \odot I
\en
The probability of obtaining the eigenvalue $\alpha_i$ in a measurement of $A_i$ at time $t_i$ is
again given by the Born formula
\eq
p(\alpha_i)= \lb \Psi | \mathbb{P}_{\alpha_i} | \Psi \rb = \sum_{\alpha_1,...,\noali~,...,\alpha_n} p(\al_1,...,\al_i,...,\al_n)
\en
This is indeed the probability of measuring $\alpha_i$ at time $t_i$, {\sl the system being measured 
also at all other times}. It is given by the sum on all sequence probabilities, keeping $\alpha_i$ fixed.

We can also consider {\sl multitime observables} $\mathbb{A}_{i,j} $, with definitions similar to
(\ref{historyobservable}), with $A_i$ and $A_j$ appearing at times $t_i$ and $t_j$, and more
generally $\mathbb{A}_{i_1,...i_m}$ where $(i_1,...,i_m)$ is a subset of $(1,2,...n)$. These
observables correspond to measurements performed at the corresponding times $t_{i_1},...t_{i_m}$.
The probability of obtaining the sequence $\alpha_{i_1},...\alpha_{i_m}$ is
\eq
p (\alpha_{i_1},...\alpha_{i_m}) = \lb\Psi | \mathbb{P}_{\alpha_{i_1},...\alpha_{i_m}} |\Psi \rb \label{prob2}
\en
where $\mathbb{P}_{\alpha_{i_1},...\alpha_{i_m}}$ is the projector on eigensubspaces of the multitime observable $\mathbb{A}_{i_1,...i_m}$:
\eq
\mathbb{P}_{\alpha_{i_1},...\alpha_{i_m}}= I \odot...\odot |\al_{i_1}\rb\lb \al_{i_1} | \odot I \odot...\odot |\al_{i_m}\rb\lb \al_{i_m} |\odot I \odot... \label{Ppartial}
\en
Again the probability (\ref{prob2}) is a sum on all sequence probabilities, keeping $\alpha_{i_1},...\alpha_{i_m}$ fixed. 

We can also compute the average values of multitime observables, i.e. the average values of the products
$\alpha_{i_1} \cdots \alpha_{i_m}$, using the familiar formula
\eq
\lb \Psi | \mathbb{A}_{i_1,...i_m} |\Psi \rb = \sum_{\alpha} p(\alpha_1,...,\alpha_n) ~\alpha_{i_1} \cdots \alpha_{i_m}
\label{averagemultitime}
\en

We have so far considered observables made out of the ``time local" observables $A_i$, that define the history Hilbert space. On this space we can consider also general observables, i.e. hermitean operators,
sum of temporal tensor products of local operators $B_i$, not necessarily commuting with the $A_i$.
\sk
\noi We first define the ``time local" observable
\eq
\mathbb{B}_i=I \odot \cdots \odot I \odot B_i \odot \cdots \odot I \label{historyobservableBi}
\en
with associated projectors on eigensubspaces:
\eq
\mathbb{P}_{\beta_i} = I \odot \cdots \odot |\beta_i\rb\lb \beta_i| \odot \cdots \odot I
\en
Now it is tempting to define the probability of obtaining a particular eigenvalue $\beta_i$ of $B_i$ in
a measurement at time $t_i$ on the system described by the history vector (\ref{historyvector}), via the Born rule
\eq
p(\beta_i) = \lb \Psi | \mathbb{P}_{\beta_i} |\Psi \rb = \sum_\alpha A^* (\alpha_1,...,\alpha_i',...,\alpha_n)
A (\alpha_1,...,\alpha_i,...,\alpha_n) \lb \alpha'_i | \beta\rb\lb \beta |\alpha_i\rb
\en
Using the identity
\eq
\sum_{ \alpha_{i+1},...,\alpha_n} A^* (\alpha_1,...,\alpha_i',...,\alpha_n) A (\alpha_1,...,\alpha_i,...,\alpha_n) = A^* (\alpha_1,...,\alpha'_i) A (\alpha_1,...,\alpha_i) \delta_{\alpha_i \alpha'_i} \label{AAidentity}
\en
we finally find
\eqa
& & p(\beta_i) = \sum_{\alpha} A^* (\alpha_1,...,\alpha_i) A (\alpha_1,...,\alpha_i) |\lb \beta_i | \alpha_i \rb|^2=
\sum_{\alpha} p(\alpha_1,...,\alpha_i)  |\lb \beta_i  | \alpha_i \rb|^2 = \nonumber \\
& & ~~~~~~ =  \sum_{\alpha} p(\alpha_1,...,\alpha_{i-1}, \beta) \label{probbeta}
\ena
This is indeed the probability of obtaining $\beta$ at time $t_i$, after having measured the observables 
$A_i$ at times $t_1,...t_{i-1}$ and then measuring $B_i$ at time $t_i$. This probability can be verified experimentally by a statistical analysis of the measurement outcomes at successive times $t_1,...,t_i$.
\sk
We can also consider multitime observables ${\mathbb{B}_{i_1 ...i_m}}$, in analogy with
${\mathbb{A}_{i_1 ...i_m}}$, and similarly compute $ \lb \Psi | \mathbb{P}_{\beta_{i_1 ...i_m}} |\Psi \rb$.
Here a problem arises: a measurement of $B_{i_1}$ produces a collapse of the system into
an eigenvector of $B_{i_1}$, and since $B_{i_1}$ does not commute in general with $A_{i_1}$, this disrupts
the history scheme based on the $A_1,...A_n$ eigenstates. No disturbance occurs in the previous case of a single local  ${B}_i$, since only  ${A}_j$ measurements antecedent to the measurement of ${B}_i$ are relevant in the expression of the probability (\ref{probbeta}). In the Appendix A we treat the case of a two-time observable, and give an operative interpretation of the Born rule in terms of an ``intermediate" vector related to the two-vector formalism of ref. \cite{twostate0,twostate1,twostate2,twostate3}. 

In Section 6 we address this question in full generality, and propose a way to interpret statistically
the scalar product $ \lb \Psi | \mathbb{P}_{\beta_{i_1 ...i_m}} |\Psi \rb$ by means of an operative protocol that transforms time evolution into compositeness.

%%%%%%%%%%%%%%%%%%%%
 \sect{Spatial and temporal entanglement entropy}
 %%%%%%%%%%%%%%%%%%%%
The history vector formalism allows for a simple definition of a history density matrix, and of mixed history states.

We first introduce a tensor product in history space, i.e. in the vector space spanned by the basis vectors 
$ |\al_1\rb \odot ... \odot |\al_n\rb$. It is defined by
\eq
 (|\al_1\rb \odot ... \odot |\al_n\rb) ( |\be_1\rb \odot ... \odot |\be_n\rb) \equiv
  |\al_1\rb |\be_1 \rb \odot ... \odot |\al_n \rb | \be_n\rb
  \en
where $ |\al_1\rb \odot ... \odot |\al_n\rb$ span the history space of a system A, and $ |\be_1\rb \odot ... \odot |\be_n\rb$
do the same for a system B. The product is then extended by bilinearity on all linear combinations of these vectors. No symbol is used 
for this tensor product, to distinguish it from the tensor product $\odot$ involving different times $t_k$.
{\sl Product history states} are then defined to be
expressible in the form:
\eq
 (\sum_\al A(\phi,\al) |\al_1\rb \odot ... \odot |\al_n\rb ) (  \sum_\be A(\chi,\be) |\be_1\rb \odot ... \odot |\be_n\rb ) \label{hproduct}
 \en
or, using bilinearity:
\eq
\sum_{\al,\be} A(\phi,\al) A(\chi,\be) |\al_1 \be_1 \rb \odot...\odot |\al_n \be_n \rb
\en
with $|\al_i \be_i \rb \equiv |\al_i \rb | \be_i \rb$ for short. A product history state is then
characterized by factorized amplitudes $A(\psi,\al,\be) =  A(\phi,\al) A(\chi,\be)$. If the history state cannot be expressed as a product, 
we define it to be {\sl history entangled}. In this case,
results of measurements on system A are correlated with those on system B and viceversa.
\sk
A system in the history state $|\Psi\rb$ can be described by the {\sl history density matrix}:
\eq
\rho = |\Psi\rb \lb \Psi |
\en
a positive operator satisfying $Tr(\rho) = 1$ (due to $\lb \Psi |\Psi \rb = 1$). A mixed history state has density matrix
\eq
\rho = \sum_i p_i  |\Psi_i\rb \lb \Psi_i |
\en
with $\sum_i p_i = 1$, and $\{ |\Psi_i \rb \}$ an ensemble of history states. Probabilities of measuring sequences $\al = \al_1,...\al_n$ in history state $\rho$ are given by the standard formula:
\eq
p(\al_1,...\al_n) = Tr(\rho ~ \mathbb{P}_\al)
\en
cf. equation (\ref{probvector}) for pure states.
\sk
\noi Consider now a system AB composed by two subsystems A and B, and devices measuring
observables $\Afat_i = A_i \otimes I$ and $\Bfat_i=I \otimes B_i$  at each $t_i$. Its history state is
\eq
|\Psi^{AB}\rb = \sum_{\al,\be} A(\psi,\al,\be) |\al_1 \be_1 \rb \odot...\odot |\al_n \be_n \rb
\label{PsiAB}
\en
where $\al_i,\be_i$ are the possible outcomes of a joint measurement at time $t_i$ of 
$\Afat_i$ and $\Bfat_i$.  The amplitudes $A(\psi,\al,\be)$ are computed using the general formula
(\ref{amplitude}), with projectors 
\eq
\Pfat_{\al_i,\be_i} = |\al_i,\be_i \rb\lb \al_i,\be_i|= |\al_i \rb\lb \al_i| \otimes |\be_i \rb\lb\be_i|
\en
corresponding to the eigenvalues $\al_i,\be_i$. The density matrix of AB is
\eqa
& & \rho^{AB} = |\Psi^{AB}\rb \lb \Psi^{AB}| = \nonumber \\
 & & ~~~~~ = \sum_{\al,\be,{\al}', {\be}'} 
A(\psi,\al,\be) A(\psi,{\al}',{\be}')^* (|\al_1 \be_1 \rb \odot...\odot |\al_n \be_n \rb )
 (\lb \al'_1 \be'_1 | \odot...\odot \lb \al'_n \be'_n |) \nonumber\\
\label{rhoAB}
\ena
We define {\sl space-reduced density matrices} by partially tracing on the subsystems:
\eq
\rho^A \equiv Tr_B  ( \rho^{AB} ), ~~~\rho^B \equiv Tr_A  ( \rho^{AB} ) \label{rhoA}
\en
In general $\rho^A$ and $\rho^B$ will not describe pure history states anymore. These reduced density matrices can be used to compute statistics for
measurement sequences on the subsystems.  Taking for example the partial trace on B of (\ref{rhoAB}) yields:
\eq
\rho^A = \sum_{\alpha,{\alpha}',\beta} A(\psi,\alpha,\beta) A^* (\psi,{\alpha}',\beta) 
(|\al_1  \rb \odot...\odot |\al_n \rb) ( \lb {\al}'_1| \odot...\odot \lb {\al}'_n  |),
\en
a positive operator with unit trace. The standard expression in terms of $\rho^A$ for Alice's probability to obtain the sequence  $\al$ is 
\eq
p(\alpha) = Tr(\rho^A \Pfat_\alpha)  \label{prho}
\en
with
\eq
\Pfat_\alpha = P_{\al_1} \odot \cdots \odot P_{\al_n} ,~~~P_{\al_i} =
 |\al_i\rb\lb\al_i| \label{Pfat}
\en
The prescription (\ref{prho}) yields
\eq
p(\alpha) =  \sum_\beta  |A(\psi,\al,\be) |^2 = \sum_\beta p(\alpha,\beta)  \label{psum}
\en
i.e. the probability for Alice to obtain the sequence $\al$ in measuring the observables $A_i$, in presence
of measurements by Bob at all times $t_i$. As observed in \cite{LC2}, if Bob does not measure the $B$ observables at all times, 
the probability for Alice is in general different from (\ref{psum}). It remains the same when the evolution operator
of the AB system factorizes ($U^{AB}=U^A \otimes U^B$) so that A and B do not interact. In this case Bob cannot
communicate with Alice by activating (or not activating) his measuring devices.
\sk
Given the $\odot$ tensor structure, we can easily define {\sl temporal entanglement}, see ref. \cite{LC3}. 
Partial traces of the history density matrix can be taken also on the Hilbert spaces ${\cal H}_i$ corresponding to different times $t_{\{k\}} = t_{k_1},...t_{k_p}$, $p<n$. We call the resulting density matrices, involving only the complementary times $t_{\{j\}} = t_{j_1},...t_{j_m}$ (i.e. with $j_1,...j_m$ and $k_1,...k_p$ having no intersection, and union coinciding with $1,...n$), {\sl time-reduced density matrices}. They 
are used to compute sequence probabilities corresponding to measurements at times $t_{\{j\}} $, given that measurements are performed also at times $t_{\{k\}}$ without registering their result. Thus they describe statistics for an experimenter that has access only to the measuring apparati at times $t_{\{j\}}$, while
the system gets measured at all times $t_i=t_1,...t_n$. For details and examples see ref. \cite{LC3}.
\sk
Finally from the von Neumann entropy $S=-Tr ( \rho \log \rho)$ can be used to characterize entanglement
between subsystems of a composite system (using the ``space" reduced $\rho$), or between Hilbert spaces of a same system considered at different times and with different measuring devices (using the ``time" reduced" $\rho$), cf. ref.s \cite{LC2,LC3}.

%%%%%%%%%%%%%%%%%%%%
 \sect{Modeling evolution with compositeness}
 %%%%%%%%%%%%%%%%%%%%
 
 In this Section we describe a protocol that replaces the temporal tensor product
 $\odot$ with the usual tensor product $\otimes$. In other words, we simulate with a
 composite system the time evolution of a single system. As we have recalled in the Introduction, this is easily done in the case of a classical system, but not so obviously done with a quantum system. 
 Some references on this issue can be found in \cite{EC1,EC2}. 
 
 We want to find an operative way to construct the composite state
 \eq
|\Psi\rb = \sum_{\alpha_1,...\alpha_n} A(\psi,\alpha_1,...,\alpha_n) ~  |\alpha_1\rb \otimes  \cdots \otimes|\alpha_n\rb \label{historyvector2}
\en
without necessarily knowing the initial state $|\psi\rb$, but having at our disposal the
eigenstates $|\alpha_i\rb$ of the $A_i$ observables. 
Moreover the evolution operator is supposed to be known, and implementable with appropriate unitary gates. Finally we will also need a gate that generalizes the CNOT 2-qubit gate. With these resources, the protocol runs as follows:
\sk
\noi 1) Evolve the initial state $|\psi\rb$ to time $t_1$. The state of the system becomes
\eq
|\psi (t_1)\rb = \sum_{\alpha_1} A(\psi,\alpha_1) |\alpha_1\rb
\en
 \noi 2) Adjoin to the system another twin system in one of the eigenstates $|\alpha_1\rb$, say 
 $|\alpha'_1\rb$:
 \eq
 \sum_{\alpha_1} A(\psi,\alpha_1) |\alpha_1\rb \otimes  |\alpha'_1 \rb  \label{state2}
 \en
\noi 3) By means of a unitary gate $V$, transform the state (\ref{state2}) into the ``diagonal state"
\eq
 \sum_{\alpha_1} A(\psi,\alpha_1) |\alpha_1\rb \otimes  |\alpha_1 \rb
 \en
 The unitary gate acts on ${\cal H}_1 \otimes {\cal H}_1$ and is defined to transform
 $ |\alpha_1\rb \otimes  |\alpha'_1 \rb$ into  $ |\alpha_1\rb \otimes  |\alpha_1 \rb$ for any
$ |\alpha_1\rb$ (and for a {\sl fixed}  $|\alpha'_1 \rb$)  , i.e.
 to clone the orthogonal states of the basis $ \{ |\alpha_1\rb \}$. If  ${\cal H}_1$ is the Hilbert space of a qubit, the unitary gate is just the CNOT gate. In general, labeling by $|\alpha^i \rb$ the i-th basis vector of ${\cal H}_1$, and choosing
 $|\alpha^1 \rb$ as the special ket $|\alpha'_1\rb$ in (\ref{state2}), the gate $V$ can be defined on the basis vectors of 
 ${\cal H}_1 \otimes {\cal H}_1$ as follows:
 \eqa
 & & V (|\alpha^i\rb \otimes |\alpha^1\rb) =  |\alpha^i\rb \otimes |\alpha^i\rb \\
 & & V (|\alpha^i\rb \otimes |\alpha^j\rb) =  |\alpha^i\rb \otimes |\alpha^j\rb ~~{\rm for~j \not=1~and~i \not= j}\\
  & & V (|\alpha^i\rb \otimes |\alpha^i\rb) =  |\alpha^i\rb \otimes |\alpha^1\rb ~~{\rm for~i \not= 1}
  \ena
 and can easily be checked to be unitary.
 \sk
\noi 4) Finally, evolve the second system to time $t_2$, leaving the first system as it is. The 
state of the bipartite system becomes
\eq
 \sum_{\alpha_1} A(\psi,\alpha_1) |\alpha_1\rb \otimes \sum_{\alpha_2} A(\alpha_1,\alpha_2)  |\alpha_2 \rb
 = \sum_{\alpha_1,\alpha_2} A(\psi,\alpha_1,\alpha_2) |\alpha_1\rb \otimes |\alpha_2 \rb
\en
after using bilinearity of the tensor product and the merging property of amplitudes $A(\psi,\alpha_1) A(\alpha_1,\alpha_2) = A(\psi,\alpha_1,\alpha_2)$.
\sk
\noi 5) Adjoin a vector $|\alpha'_2\rb$. Repeat the procedures 3) and 4) for the states  $ |\alpha_2\rb \otimes  |\alpha'_2 \rb$. Cycle until the state (\ref{historyvector2}) is reached. 

This protocol permits to investigate the temporal characteristics of a quantum system, as for example
temporal entanglement between states of the system at different times, via measurements on a static composite system,
which are usually more easily performed.

%%%%%%%%%%%%%%%%%%%%%%%%%%%%%
\sect{LG inequality is never violated in consistent histories}
%%%%%%%%%%%%%%%%%%%%%%%%%%%%%

The Leggett-Garg (LG) inequality, proposed in ref. \cite{LG}, is derived within the context of {\sl macrorealism}, based on two fundamental assumptions:
\sk
\noi 1) a physical system with two or more distinct states available to it will at all times be in one or the other of these states.

\noi 2) it is possible, in principle, to determine the state of the system with arbitrarily small perturbation on its subsequent dynamics. 
\sk
\noi A third assumption may be added:

\noi 3) the result of a measurement remains unaffected by any future choices regarding what will or will not be measured.
\sk
The latter assumption, together with the second, form the general assumption of {\sl locality in time}, with the first taking the role of {\sl realism}. In a probabilistic theory (i.e. a theory that assigns probabilities to measurement outcomes), the three above assumptions imply the classical marginal rules (\ref{sumrulep}).

Consider now a system with a dichotomic variable $Q$, which can take $q= \pm 1$ as values. Its
two-time correlation function is defined by
\eq
C_{ij} = \sum_{q_i,q_j = \pm 1} q_i q_j ~p(q_i, q_j)
\en
where $p(q_i,q_j)$ is the joint probability of obtaining $q_i$ and $q_j$ as results of measurements at times $t_i$ and $t_j$ respectively. Under the assumptions of macrorealism (i.e. using the classical marginal rules
for probabilities), and considering three times $t_1,t_2,t_3$, we have:
\eqa
& & C_{12}= \sum_{q=\pm 1} [p(q,q,\ast) - p(q,-q,\ast)] \nonumber \\
& & C_{13}= \sum_{q=\pm 1} [p(q,\ast,q) - p(q,\ast,-q)] \nonumber \\
& & C_{23}= \sum_{q=\pm 1} [p(\ast,q,q) - p(\ast,q,-q)] 
\ena
where $p(q,q,\ast) \equiv p(q,q,+1)+p(q,q,-1)$ is the joint probability of measuring $q$ at time $t_1$ and the same $q$ at time $t_2$, etc.  Then the following equality is easily derived:
\eq
C_{12}+C_{13}-C_{23} = 1 - 4 ~[p(+1,-1,+1) + p(-1,+1,-1)]
\en
This quantity has $+1$ as upper bound and $-3$ as lower bound, and we have therefore the Leggett-Garg inequality \cite{LG}:
\eq
-3 \leq C_{12}+C_{13}-C_{23} \leq 1
\en
This inequality is violated in various quantum systems, signalling that at least one of the assumptions used to derive it does not hold in the quantum world.

We now demostrate that the inequality is never violated when considering only 
consistent histories, i.e. histories satisfying condition (\ref{decocondition}). Consider a quantum
system starting in the initial state $|\psi\rb$, with devices measuring the same dichotomic observable $Q$ at times $t_1,t_2,t_3$. There are eight different histories $(q_1,q_2,q_3)$ with $q_i=\pm 1$, with
probabilities:
\eq
p(q_1,q_2,q_3) = Tr ( C_{\psi,q_1,q_2,q_3} C^\dagger_{\psi,q_1,q_2,q_3} )
\en
where
\eq
C_{\psi,q_1,q_2,q_3}  \equiv P_{q_3} U_{23} P_{q_2} U_{12} P_{q_1} U_{01} |\psi\rb \lb \psi |
\en
is the chain operator and $U_{ij}$ is the evolution operator between $t_i$ and $t_j$. Now the classical marginal rules do not hold in general: for example the joint probability
of obtaining $q_2$ at $t_2$ and $q_3$ at $t_3$, which we still denote by $p(\ast,q_2,q_3)$, is
given by
\eq
p(\ast,q_2,q_3) = p(+1,q_2,q_3) + p(-1,q_2,q_3) + {\cal I} (\ast,q_2,q_3)
\en
where 
\eq
{\cal I} (\ast,q_2,q_3) = 2 ~Re ~Tr [ C_{\psi,+1,q_2,q_3} C^\dagger_{\psi,-1,q_2,q_3} ] 
\en
The LG inequality takes the form:
\eq
-3 \leq 1 - \sum_q  [ 4p(q,-q.q) - {\cal I} (\ast,q,q) + {\cal I} (\ast,q,-q)+ {\cal I} (q, \ast,q)- {\cal I} (q, \ast,-q) ]\leq 1
\en
where the two interference terms ${\cal I} (q,q,\ast), {\cal I} (q,-q,\ast)$ are absent because they vanish,
since in this case the classical sum rule (\ref{sumrulep}) holds.  We see that for the violation of LG, at least one of the four interference terms must be nonvanishing, implying at least a pair of inconsistent histories. 

Thus a formalism that considers only consistent histories will never reveal any violation of the LG inequality. This agrees 
with an analogous conclusion in ref. \cite{PM}.

%%%%%%%%%%%%%%%%%%%%%%%%
\sect{Temporal CHSH inequality }
%%%%%%%%%%%%%%%%%%%%%%%%

Consider an observer, Alice, and allow her to perform a measurement at time $t_1$ choosing
between two dichotomic variables $A_1, B_1$, and then, at $t_2$, to measure choosing between
two dichotomic variables $A_2, B_2$. In analogy with the spatial CHSH inequality \cite{CHSH}, we define the quantity
\eq
A_1 (A_2+B_2) + B_1 (A_2-B_2) = \pm 2
\en
By averaging this expression over multiple runs of the measurement sequence, we can derive the following temporal CHSH inequality
\eq
\lb A_1 A_2 \rb + \lb A_1 B_2\rb + \lb B_1 A_2\rb - \lb B_1 B_2 \rb \leq 2 \label{tCHSH}
\en
As it is the case for the spatial CHSH inequality, also its temporal analogue can be violated in quantum systems \cite{BTCV}. In the history vector formalism we could just repeat the algebra leading to the spatial CHSH inequality violation, since the tensor structures are identical (replacing $\otimes$ with $\odot$ tensor products). For example using the two-time entangled history state
\eq
|\Psi\rb = {1 \over \sqrt{2}} (|1\rb \odot |0\rb - |0\rb \odot |1\rb) \label{bell2}
\en
and choosing $A_1=X,B_1=Z, A_2 = -{Z+X \over \sqrt{2}}, B_2 = {Z-X \over \sqrt{2}}$, we find
the expectation values:
\eqa
& & \lb A_1 \odot A_2 \rb_\Psi = \lb \Psi | X \odot \left( -{Z+X \over \sqrt{2}}\right) |\Psi \rb = {1\over \sqrt{2}} \\
& & \lb B_1 \odot A_2 \rb_\Psi = \lb \Psi | Z \odot \left( -{Z+X \over \sqrt{2}}\right) |\Psi \rb = {1\over \sqrt{2}} \\
& & \lb A_1 \odot B_2 \rb_\Psi = \lb \Psi | X \odot \left( {Z-X \over \sqrt{2}}\right) |\Psi \rb = {1\over \sqrt{2}} \\
& & \lb B_1 \odot B_2 \rb_\Psi = \lb \Psi | Z \odot \left( {Z-X \over \sqrt{2}}\right) |\Psi \rb = - {1\over \sqrt{2}} 
\ena
explicitly violating (\ref{tCHSH}). 

However, in order to check experimentally this violation one has to use the procedure of Appendix A
for a measurement of a two-time observable, or to use the general protocol of Section 5 that simulates
 the evolution in time of a quantum system with a composite system at a fixed time. Note that in computing 
 the above averages we have used in all four cases the same history vector (\ref{bell2}). This procedure gives averages related to measurements in the computational basis (for example measurements of the $Z$ observable),
 followed by measurements of the $A,B$ observables. They are really ``watermarked" by the choice of the computational basis. To find the averages of $A,B$ measurements at $t_1$ and $t_2$, without reference to the computational basis (i.e. with only $A,B$ devices activated), we have to choose the correct history vector for each of the four cases, corresponding to the observables $A,B$ involved. This we do in the next paragraph. We do not expect the same averages as obtained above, but 
 the differences in this case will just amount to signs, and by redefining $A_1 \rightarrow -A_1$ we recover 
  the violation of the temporal CHSH inequality.
  \sk
 The history vector (\ref{bell2}) can be obtained from the initial state at $t_0$:
 \eq
 |\psi\rb = {1 \over \sqrt{2}} (|1\rb - |0\rb)
 \en
 where $|0\rb$ and $|1\rb$ are eigenvectors of $Z$ with eigenvalues $+1$ and $-1$ respectively, and with evolution operators given by $U(t_1,t_0)=I$, $U(t_2,t_1)=X$. The history state of the system, if we choose $Z$ as the measured observable at $t_1$ and $t_2$, is the temporal analogue of a two-qubit Bell state given in (\ref{bell2}). 
 
 To compute average values of the observables $A_1 \odot A_2$ etc., i .e. the average of the products $a_1 a_2$ where $a_1$ is the result of a measurement of $A_1$ at time $t_1$ (and similar for $a_2$), we need the history vectors $|\Psi\rb$ corresponding to the particular choice of observables $A_1, A_2$, differing from
 the history vector corresponding to $Z$ measurements. Using these history vectors, the average of 
 $A_1 \odot A_2$ can be computed as
 \eq
 \lb \Psi | A_1 \odot A_2 |\Psi \rb = \sum_{a_1=\pm1,a_2=\pm 1}  p(a_1,a_2)  a_1 a_2
 \en
cf. equation (\ref{averagemultitime}). The probabilities $p(a_1,a_2)$ are given by the square modulus of the amplitudes 
\eq
A(\psi, a_1,a_2) = \lb a_2 | X P_{a_1} |\psi \rb
\en
according to formula (\ref{amplitude}). We can list the probabilities for all the histories 
in each of the four cases ($A_1,A_2$), ($B_1,A_2$), ($A_1,B_2$), ($B_1,B_2$):
\eqa
& & (A_1,A_2): ~~~p(+1,\pm 1)=0,~~p(-1,\pm 1)= {1 \over 4 \mp \sqrt{2}} \\
& & (B_1,A_2): ~~~p(+1,\pm 1)= {1 \over 2(4 \mp \sqrt{2})} ,~~p(-1,\pm 1)= {3\mp 2\sqrt{2} \over 2(4 \mp \sqrt{2})} \\
& & (A_1,B_2): ~~~p(+1,\pm 1)= 0 ,~~p(-1,\pm 1)= {3\pm 2\sqrt{2} \over 4 \pm \sqrt{2}} \\
& & (B_1,B_2): ~~~p(+1,\pm 1)= {1 \over 2(4 \pm \sqrt{2})} ,~~p(-1,\pm 1)= {3\pm 2\sqrt{2} \over 2(4 \pm \sqrt{2})} \\
\ena
The averages are then given by:
\eqa
& & \lb A_1 \odot A_2 \rb = p(-1,+1) ~(-1)(+1) + p(-1,-1)~(-1)(-1)  = -{1 \over 4 - \sqrt{2}} +{1 \over 4 + \sqrt{2}} = -{1\over \sqrt{2}}\nonumber \\
& &  \lb B_1 \odot A_2 \rb =  {1 \over 2(4 -\sqrt{2})}  - {1 \over 2(4 + \sqrt{2})}  -   {3- 2\sqrt{2} \over 2(4 - \sqrt{2})}
+ {3 + 2\sqrt{2} \over 2(4 + \sqrt{2})} = + {1\over \sqrt{2}}\nonumber\\
& & \lb A_1 \odot B_2 \rb = - {3 + 2\sqrt{2} \over 4 + \sqrt{2}} +  {3 - 2\sqrt{2} \over 4 - \sqrt{2}}= -{1\over \sqrt{2}}\nonumber \\
& &  \lb B_1 \odot B_2 \rb =  {1 \over 2(4 + \sqrt{2})}  - {1 \over 2(4 - \sqrt{2})}  -   {3 + 2\sqrt{2} \over 2(4 + \sqrt{2})}
+ {3 - 2\sqrt{2} \over 2(4 - \sqrt{2})} = - {1\over \sqrt{2}}\nonumber\\
\ena
Taking now $A_1 = -X$ instead of $A_1 =X$, the averages above become respectively $+{1\over \sqrt{2}},+{1\over \sqrt{2}},+{1\over \sqrt{2}},-{1\over \sqrt{2}}$ and violate the temporal CHSH inequality (\ref{tCHSH}).

\sect{Conclusions}

We have explored some selected applications of the history vector formalism. Advantages of this
approach are:
\sk
\noi $\bullet$ it is a natural generalization of the state vector expansion on a basis of observable eigenvectors;

\noi $\bullet$ the formalism incorporates measuring devices in the description of the state evolution. Different
devices imply different history states, even if initial state and evolution operator coincide.

\noi $\bullet$ space and time correlations can be computed with similar algebraic operations;

\noi $\bullet$ the similarity of time and space tensor products permits to map an evolving system into 
a static composite system;

\noi $\bullet$ there is no need to consider only consistent histories.

\section*{Acknowledgements}

This work is supported by the research funds of the Eastern Piedmont University and INFN - Torino Section. 

\app{Appendix: two-time observable}

Consider the (two-time) history vector 
\eq
|\Psi\rb = \sum_\alpha A(\psi,\alpha_1,\alpha_2) |\alpha_1 \rb \odot |\alpha_2 \rb 
\en
and the two-time observable
\eq
\mathbb{B}= B_1 \odot  B_2  \label{historyobservableB12}
\en
with associated projectors on eigensubspaces
\eq
\mathbb{P}_{\beta_1,\beta_2} =  |\beta_1\rb\lb \beta_1| \odot  |\beta_2\rb\lb \beta_2|  \label{projector12}
\en
Computing the expectation value of the projector (\ref{projector12}) yields:
\eqa
& & p(\beta_1,\beta_2) = \lb \Psi | \mathbb{P}_{\beta_1,\beta_2} |\Psi \rb =
 \sum_{\alpha,\alpha'} A^* (\psi,\alpha'_1,\alpha'_2)
A (\psi,\alpha_1,\alpha_2) \lb \alpha'_1 | \beta_1\rb\lb \beta_1 |\alpha_1\rb
 \lb \alpha'_2 | \beta_2\rb\lb \beta_2 |\alpha_2\rb \nonumber \\
 & & =  \sum_{\alpha,\alpha'} A^* (\psi,\alpha'_1,\beta_2)
A (\psi,\alpha_1,\beta_2) \lb \alpha'_1 | \beta_1\rb\lb \beta_1 |\alpha_1\rb
\ena
We can rewrite this joint probability as a one-time expectation value of the projector $P_{\beta_1}$
\eq
 p(\beta_1,\beta_2)  = \lb \psi_1|P_{\beta_1} |\psi_1 \rb
 \en
 where the ``intermediate state" $ |\psi_1 \rb$ is defined as the (normalized) superposition at time $t_1$
 \eq
  |\psi_1 \rb = {1 \over \sqrt{{\cal N}}} \sum_{\alpha_1} A(\psi,\alpha_1,\beta_2) |\alpha_1 \rb,~~~{\cal N}=\sum_{\alpha_1} |A(\psi,\alpha_1,\beta_2)|^2 = \sum_{\alpha_1} p(\psi,\alpha_1,\beta_2)
  \en
  This intermediate state is the state at $t_1$, depending on the preselected initial state $|\psi \rb$ and on the postselected 
  final state $|\beta_2\rb$, for which a measurement of $A_1$ has a probability
 $|A(\psi,\alpha_1,\beta_2)|^2/{\cal N}$ to yield $\alpha_1$. It is related to the two-state vector formalism of Aharonov, Bergmann and Lebowitz (ABL): indeed its squared amplitudes reproduce the ABL rule for intermediate probabilities of obtaining $\alpha_1$, given that the state at $t_0$ is $|\psi\rb$ and at $t_2$ is $|\beta_2\rb$.

\vfill\eject
\end{document}